\documentclass[doublecol]{epl2}
\usepackage{amssymb}
\usepackage{amsmath}
\usepackage{graphicx}
\usepackage{epstopdf}
\usepackage{dcolumn}
\usepackage{bm}
\usepackage{amsfonts}%
\title{Damping in a Superconducting Mechanical Resonator}
\author{Oren Suchoi\inst{1} \and Eyal Buks\inst{1}}
\shortauthor{F. Author \etal}

\institute{
  \inst{1} Andrew and Erna Viterbi Department of Electrical Engineering, Technion, Haifa 32000 Israel\\
}

\pacs{74.25.Ld}{Superconductivity , Mechanical and acoustical properties, elasticity, and ultrasonic attenuation}
\pacs{46.40.- f}{vibrations and mechanical waves}

\abstract{
We study a mechanical resonator made of aluminum near the normal to super
conductivity phase transition. A sharp drop in the rate of mechanical damping
is observed below the critical temperature. The experimental results are
compared with predictions based on the Bardeen Cooper Schrieffer theory of
superconductivity and a fair agreement is obtained.
}

\begin{document}
\maketitle



%It is always \today, today,
%but any date may be explicitly specified

%PACS, the Physics and Astronomy
%Classification Scheme.

Mechanical resonators having low damping rate are widely employed for sensing
and timing applications \cite{Ekinci_061101}. At sufficiently low temperatures
such devices may allow the experimental exploration of the crossover from
classical to quantum mechanics
\cite{Meystre2013,Poot_273,OConnell_697,Pikovski_393,Weinstein_041003,Teufel_359,Chan_89}%
. Commonly, the observation of non-classical effects in such experiments is
possible only when the damping rate \cite{Unterreithmeier_027205} of the
mechanical resonator is sufficiently low. Mechanical resonators made of
superconductors are widely employed in such low-temperature experiments. In
this study we experimentally investigate the effect of superconductivity on
the damping rate of a mechanical resonator made of aluminum near its normal to
super phase transition.

The damping rate can be measured by coupling the mechanical resonator under
study to a displacement detector. In general, a variety of different
mechanisms may contribute to the total mechanical damping rate. In order to
isolate the effect of superconductivity it is important to employ a method of
displacement detection that is unaffected by the normal to super phase
transition. In addition, systematic errors in the measured damping rate due to
back-reaction effects originating from the coupling between the mechanical
resonator and the displacement detector have to be kept at a sufficiently low level.%

%TCIMACRO{\FRAME{ftbpFU}{3.4537in}{2.5949in}{0pt}{\Qcb{The experimental setup.
%(a) A sketch of the mechanical resonator and the fiber-based optical cavity.
%(b) Electron micrograph of the trampoline. (c) The resonance lineshape of the
%fundamental mechanical mode.}}{\Qlb{Fig setup}}{fig_setup.eps}%
%maintain-aspect-ratio TRUE;  display "ICON";  valid_file "F";
%width 3.4537in;  height 2.5949in;  depth 0pt;  original-width 11.3805in;
%original-height 8.5282in;  cropleft "0";  croptop "1";  cropright "1";
%cropbottom "0";  filename 'Fig_setup.eps';file-properties "XNPEU";}}}%
%BeginExpansion
\begin{figure}
[ptb]
\begin{center}
\includegraphics[
height=2.5949in,
width=3.4537in
]%
{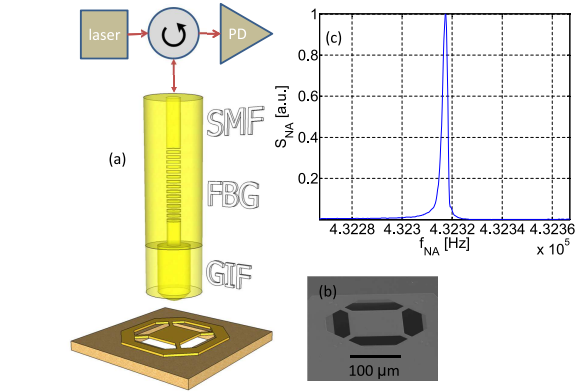}%
\caption{The experimental setup. (a) A sketch of the mechanical resonator and
the fiber-based optical cavity. (b) Electron micrograph of the trampoline. (c)
The resonance lineshape of the fundamental mechanical mode.}%
\label{Fig setup}%
\end{center}
\end{figure}
%EndExpansion

In our setup (see Fig. \ref{Fig setup}) we employ the so-called optomechanical
cavity configuration
\cite{Braginsky&Manukin_67,Aspelmeyer_1391,Metcalfe_031105}, in which
displacement detection is performed by coupling the mechanical resonator to an
electromagnetic cavity. While many of the previous studies of superconducting
mechanical resonators have employed such a configuration with a
superconducting microwave cavity
\cite{Teufel_204,Teufel_359,OConnell_697,Palomaki_710,Weinstein_041003,Yanai_1608_01871,Andrews_10021,Lecocq_043601,Suh_6260,Wollman_952,Pirkkalainen_243601,Suchoi_043829}%
, our setup, which is based on a cavity in the optical band, allows
displacement detection that is unaffected by the phase transition occurring in
the mechanical resonator under study, which, in-turn, allows isolating the
effect of superconductivity on the mechanical damping. Moreover, back-reaction
effects are suppressed by employing a relatively low driving power to the
optical cavity (see discussion below).

In our setup the optomechanical cavity is formed between two mirrors, a
stationary fiber Bragg grating (FBG) mirror and a movable mirror made of a
mechanical resonator in the shape of a trampoline supported by four beams [see
Fig. \ref{Fig setup}(a)]. A graded index fiber (GIF) spliced to the end of the
single mode fiber (SMF) is employed for focusing. A cryogenic piezoelectric
three-axis positioning system having sub-nanometer resolution is employed for
manipulating the position of the optical fiber. A photo-lithography process is
used to pattern a $t_{\mathrm{Al}}=200%
%TCIMACRO{\unit{nm}}%
%BeginExpansion
\operatorname{nm}%
%EndExpansion
$ thick aluminum layer on top of a a $t_{\mathrm{SiN}}=100%
%TCIMACRO{\unit{nm}}%
%BeginExpansion
\operatorname{nm}%
%EndExpansion
$ thick silicon nitride layer into the shape of a $100\times100%
%TCIMACRO{\unit{\U{3bc}m}}%
%BeginExpansion
\operatorname{\mu m}%
%EndExpansion
^{2}$ trampoline with four supporting beams [see Fig. \ref{Fig setup}(b)].
Details of the fabrication process can be found elsewhere \cite{Suchoi_033818}%
. Measurements are performed in a dilution refrigerator at a pressure well
below $2\times10^{-5}%
%TCIMACRO{\unit{mbar}}%
%BeginExpansion
\operatorname{mbar}%
%EndExpansion
$. Another device on the same wafer, which is made in the shape of a microwave
microstrip resonator, allows characterizing the surface resistance of the
aluminum layer \cite{Suchoi_033818}.

A tunable laser operating near the Bragg wavelength of the FBG together with
an external attenuator are employed to excite the optical cavity. The optical
power reflected off the cavity is measured by a photodetector (PD), which is
connected to a network analyzer (NA). Actuation is performed by applying an
alternating voltage (with a direct voltage offset) between the trampoline and
a stationary electrode positioned $200%
%TCIMACRO{\unit{\U{3bc}m}}%
%BeginExpansion
\operatorname{\mu m}%
%EndExpansion
$ below it.

The fundamental mechanical mode is characterized by its frequency
$\omega_{\mathrm{m}}/2\pi$ and damping rate $\gamma_{\mathrm{m}}$. Both
parameters can be extracted from the resonance lineshape of the measured NA
signal $S_{\mathrm{NA}}$ vs. angular driving frequency $\omega_{\mathrm{NA}}$
with a fixed driving amplitude [see Fig. \ref{Fig setup}(c)]. In the regime of
linear response $S_{\mathrm{NA}}\left(  \omega_{\mathrm{NA}}\right)  $ is
expected to be given by%
\begin{equation}
S_{\mathrm{NA}}\left(  \omega_{\mathrm{NA}}\right)  =\frac{S_{\mathrm{NA,R}}%
}{1+\left(  \frac{\omega_{\mathrm{NA}}-\omega_{\mathrm{m}}}{\gamma
_{\mathrm{m}}}\right)  ^{2}}{}, \label{S_NA}%
\end{equation}
where $S_{\mathrm{NA,R}}$ is the value of $S_{\mathrm{NA}}\left(
\omega_{\mathrm{NA}}\right)  $ at resonance, i.e. when $\omega_{\mathrm{NA}%
}=\omega_{\mathrm{m}}$.

\begin{figure}
[ptb]
\begin{center}
\includegraphics[
height=2.6595in,
width=3.2397in
]%
{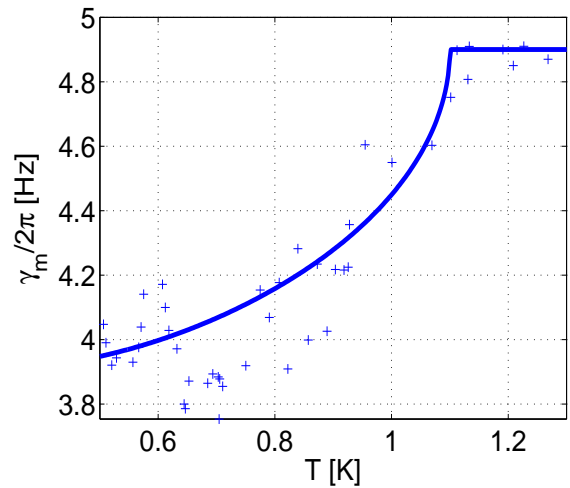}%
\caption{The measured (crosses) and theoretically calculated (solid line)
damping rate $\gamma_{\mathrm{m}}$ vs. temperature $T$. The calculated
$\gamma_{\mathrm{m}}$ is obtained using Eq. (\ref{gamma_m BCS}).}%
\label{Fig_gamma}%
\end{center}
\end{figure}

The measured damping rate $\gamma_{\mathrm{m}}$ vs. temperature $T$, which is
extracted from the NA data using Eq. (\ref{S_NA}), is indicated by the crosses
in Fig. \ref{Fig_gamma}. The same procedure yields the mode's frequency
$\omega_{\mathrm{m}}/2\pi=432.318%
%TCIMACRO{\unit{kHz}}%
%BeginExpansion
\operatorname{kHz}%
%EndExpansion
$, which is found to be almost a constant for the range of temperatures
explored in this measurement (between $0.5%
%TCIMACRO{\unit{K}}%
%BeginExpansion
\operatorname{K}%
%EndExpansion
$ and $1.3%
%TCIMACRO{\unit{K}}%
%BeginExpansion
\operatorname{K}%
%EndExpansion
$). As can be seen from Fig. \ref{Fig_gamma}, the measured damping rate
$\gamma_{\mathrm{m}}$ sharply drops as the temperature $T$ is lowered below
the value of $1.1%
%TCIMACRO{\unit{K}}%
%BeginExpansion
\operatorname{K}%
%EndExpansion
$. Note that at the same temperature of $1.1%
%TCIMACRO{\unit{K}}%
%BeginExpansion
\operatorname{K}%
%EndExpansion
$ the resonance of the microwave microstrip resonator becomes visible,
indicating thus that the critical temperature is $T_{\mathrm{c}}=1.1%
%TCIMACRO{\unit{K}}%
%BeginExpansion
\operatorname{K}%
%EndExpansion
$, as is expected from high quality aluminum layer. The data seen in Fig.
\ref{Fig_gamma} is obtained by setting the laser power that is injected into
the optical cavity to the value $P_{\mathrm{L}}=3\times10^{-8}%
%TCIMACRO{\unit{W}}%
%BeginExpansion
\operatorname{W}%
%EndExpansion
$.%

%TCIMACRO{\FRAME{ftbpFU}{3.2397in}{2.6595in}{0pt}{\Qcb{The measured (crosses)
%and theoretically calculated (solid line) damping rate $\gamma_{\mathrm{m}}$
%vs. temperature $T$. The calculated $\gamma_{\mathrm{m}}$ is obtained using
%Eq. (\ref{gamma_m BCS}).}}{\Qlb{Fig_gamma}}{fig_gamma.eps}%
%maintain-aspect-ratio TRUE;  display "ICON";  valid_file "F";
%width 3.2397in;  height 2.6595in;  depth 0pt;  original-width 9.8805in;
%original-height 8.0947in;  cropleft "0";  croptop "1";  cropright "1";
%cropbottom "0";  filename 'Fig_gamma.eps';file-properties "XNPEU";}}}%
%BeginExpansion
%EndExpansion

In general, heating due to optical absorption by the aluminum layer may give
rise to a systematic error in the measurement of $\gamma_{\mathrm{m}}$. Two
possible mechanisms are discussed below. The first one is due to back-reaction
originating from the bolometric optomechanical coupling
\cite{Metzger_133903,Zaitsev_46605}, which gives rise to a shift in the
effective value of $\gamma_{\mathrm{m}}$ denoted by $\gamma_{\mathrm{m,ba}}$.
The magnitude of $\gamma_{\mathrm{m,ba}}$ can be roughly estimated using the
relation $\left\vert \gamma_{\mathrm{m,ba}}\right\vert /\gamma_{\mathrm{m}%
}\simeq P_{\mathrm{L}}/P_{\mathrm{LT}}$, where $P_{\mathrm{L}}$ is the laser
power that is employed for the measurement of $\gamma_{\mathrm{m}}$ and
$P_{\mathrm{LT}}$ is the laser power at the threshold of self-excited
oscillation \cite{Yuvaraj_210403}. For the same cavity tuning, for which the
data seen in Fig. \ref{Fig_gamma} is obtained, self-excited oscillation occurs
at a threshold power given by $P_{\mathrm{LT}}=4\times10^{5}P_{\mathrm{L}}$,
and thus the effect of back-reaction can be safely disregarded.

The other possible source of a systematic error originates from temperature
rise due to optical absorption. Due to the low heat conductance of both
superconducting aluminum and silicon nitride, this mechanism imposes a severe
upper limit upon the allowed values of laser power. The heating power is given
by $P_{\mathrm{H}}=\zeta\beta_{\mathrm{F}}\left(  1-R_{\mathrm{C}}\right)
P_{\mathrm{L}}$, where $\zeta$ is the absorption coefficient (for aluminum
$\zeta=0.03$), $\beta_{\mathrm{F}}$ is the cavity finesse, $R_{\mathrm{C}}$ is
the cavity reflectivity, and $P_{\mathrm{L}}$ is the laser power
\cite{Shlomi_032910}. For the measurement of $\gamma_{\mathrm{m}}$ that is
presented in Fig. \ref{Fig_gamma} the optical cavity is tuned to have finesse
of $\beta_{\mathrm{F}}=1.8$ and reflectivity of $R_{\mathrm{C}}=0.32$, and
thus for this measurement $P_{\mathrm{H}}=1.1\times10^{-9}%
%TCIMACRO{\unit{W}}%
%BeginExpansion
\operatorname{W}%
%EndExpansion
$.

The temperature rise $\Delta T$ due to optical absorption is estimated by
$\Delta T=P_{\mathrm{H}}/4K_{\mathrm{b}}$, where $K_{\mathrm{b}}$ is the
thermal conductance of each of the four nominally identical beams that support
the trampoline. The thermal conductance $K_{\mathrm{b}}$ is given by
$K_{\mathrm{b}}=\left(  t_{\mathrm{Al}}\kappa_{\mathrm{Al}}+t_{\mathrm{SiN}%
}\kappa_{\mathrm{SiN}}\right)  \left(  w_{\mathrm{b}}/l_{\mathrm{b}}\right)
$, where $\kappa_{\mathrm{Al}}$ ($\kappa_{\mathrm{SiN}}$) is the thermal
conductivity of aluminum (silicon nitride), and where $w_{\mathrm{b}%
}/l_{\mathrm{b}}=0.5$ is the ratio between width and length of the beams. For
the lowest value of $0.5%
%TCIMACRO{\unit{K}}%
%BeginExpansion
\operatorname{K}%
%EndExpansion
$ in the temperature range seen in Fig. \ref{Fig_gamma} the thermal
conductivities are estimated to be $\kappa_{\mathrm{Al}}\simeq4\times10^{-2}%
%TCIMACRO{\unit{W}}%
%BeginExpansion
\operatorname{W}%
%EndExpansion%
%TCIMACRO{\unit{K}}%
%BeginExpansion
\operatorname{K}%
%EndExpansion
^{-1}%
%TCIMACRO{\unit{m}}%
%BeginExpansion
\operatorname{m}%
%EndExpansion
^{-1}$ for aluminum \cite{Feshchenko_1609_06519} and $\kappa_{\mathrm{SiN}%
}\simeq10^{-2}%
%TCIMACRO{\unit{W}}%
%BeginExpansion
\operatorname{W}%
%EndExpansion%
%TCIMACRO{\unit{K}}%
%BeginExpansion
\operatorname{K}%
%EndExpansion
^{-1}%
%TCIMACRO{\unit{m}}%
%BeginExpansion
\operatorname{m}%
%EndExpansion
^{-1}$ for silicon nitride \cite{Zink_199,Woodcraft_1968}. For these values
the estimated temperature rise is $\Delta T=0.06%
%TCIMACRO{\unit{K}}%
%BeginExpansion
\operatorname{K}%
%EndExpansion
$. For temperatures below $0.5%
%TCIMACRO{\unit{K}}%
%BeginExpansion
\operatorname{K}%
%EndExpansion
$ the temperature rise $\Delta T$ becomes even larger since both
$\kappa_{\mathrm{Al}}$ and $\kappa_{\mathrm{SiN}}$ rapidly drops at low
temperatures, and therefore no reliable measurements can be obtained unless
$P_{\mathrm{H}}$ is further reduced. However, no significant reduction of
$P_{\mathrm{H}}$ is possible in our setup due to noise, and consequently
reliable data far below $0.5%
%TCIMACRO{\unit{K}}%
%BeginExpansion
\operatorname{K}%
%EndExpansion
$ cannot be obtained. Note, however, that above $0.5%
%TCIMACRO{\unit{K}}%
%BeginExpansion
\operatorname{K}%
%EndExpansion
$ no significant change in the damping rate is obtained when the measurements
are repeated with a laser power two times higher, verifying thus that heating
does not give rise to a significant systematic error in that range.

To account for the experimental results the measured damping rate
$\gamma_{\mathrm{m}}$ is compared with theory. The solid line seen in Fig.
\ref{Fig_gamma} represents the calculated value of $\gamma_{\mathrm{m}}$
obtained from the following expression \cite{Persson_145,Claiborne_A893}%
\begin{equation}
\gamma_{\mathrm{m}}=\gamma_{\mathrm{S}}+\frac{2\gamma_{\mathrm{N}}}%
{1+\exp\frac{\Delta\left(  T\right)  }{k_{\mathrm{B}}T}}{},
\label{gamma_m BCS}%
\end{equation}
where the fitting parameters $\gamma_{\mathrm{S}}$, $\gamma_{\mathrm{N}}$ and
$T_{\mathrm{c}}$\ are taken to be given by $\gamma_{\mathrm{S}}/2\pi=3.9%
%TCIMACRO{\unit{Hz}}%
%BeginExpansion
\operatorname{Hz}%
%EndExpansion
$, $\gamma_{\mathrm{N}}/2\pi=1.0%
%TCIMACRO{\unit{Hz}}%
%BeginExpansion
\operatorname{Hz}%
%EndExpansion
$ and $T_{\mathrm{c}}=1.1%
%TCIMACRO{\unit{K}}%
%BeginExpansion
\operatorname{K}%
%EndExpansion
$. The temperature dependent energy gap $\Delta\left(  T\right)  $ is found by
numerically solving the Bardeen Cooper Schrieffer (BCS) gap equation
\cite{Bardeen_1175}%
\begin{equation}
\nu=\int_{0}^{\frac{e^{\nu}}{2\delta}}\mathrm{d}x\;\frac{\tanh\left(
\frac{\xi\delta\sqrt{1+x^{2}}}{\tau}\right)  }{\sqrt{1+x^{2}}}\;,
\end{equation}
where $\nu=1/gD_{0}$ is the inverse interaction strength with $g$ being the
electron-phonon coupling coefficient and $D_{0}$ being the density of states
per unit volume, $\delta=\Delta/\Delta_{0}$ is the normalized gap with
$\Delta_{0}$ being the zero temperature gap, $\tau=T/T_{\mathrm{c}}$ is the
normalized temperature, and the number $\xi$ is given by $\xi=\pi
/2e^{C_{\mathrm{E}}}$ with $C_{\mathrm{E}}\simeq0.577$ being the Euler's
constant. As can be seen from Fig. \ref{Fig_gamma}, fair agreement between
data and theory is obtained. A similar theoretical approach has been employed
before to successfully account for the results of a measurement of a
contact-less friction between a superconducting niobium film and a cantilever
\cite{Kisiel_119} (see also \cite{Dayo_1690}).

The rate $\gamma_{\mathrm{N}}$ represents the contribution of normal electrons
to the total rate of mechanical damping [see Eq. (\ref{gamma_m BCS})]. This
rate has been calculated in Ref. \cite{Lindenfeld_085448} for the case where
the electron mean-free path is greater than the wavelength of the oscillating
acoustic mode. However, this assumption is not valid for our device. When the
electron mean-free path is shorter than the acoustic wavelength the rate
$\gamma_{\mathrm{N}}$ can be roughly estimated using Stokes' law of sound
attenuation \cite{Stokes_287,Kittel_205}, which relates this rate to the
electronic viscosity \cite{Heinisch_76}. For the case of an acoustic wave
having angular frequency $\omega_{\mathrm{m}}$ propagating in a bulk aluminum
the rate according to this approach is given by%
\begin{equation}
\gamma_{\mathrm{N,bulk}}=\frac{2\eta_{\mathrm{Al}}\omega_{\mathrm{m}}^{2}%
}{3\rho_{\mathrm{Al}}c_{\mathrm{Al}}^{2}}{},
\end{equation}
where $\rho_{\mathrm{Al}}=2.7%
%TCIMACRO{\unit{g}}%
%BeginExpansion
\operatorname{g}%
%EndExpansion%
%TCIMACRO{\unit{cm}}%
%BeginExpansion
\operatorname{cm}%
%EndExpansion
^{-3}$ and $c_{\mathrm{Al}}=5.1\times10^{3}%
%TCIMACRO{\unit{m}}%
%BeginExpansion
\operatorname{m}%
%EndExpansion%
%TCIMACRO{\unit{s}}%
%BeginExpansion
\operatorname{s}%
%EndExpansion
^{-1}$ are the mass density and the speed of sound, respectively, of aluminum,
and where $\eta_{\mathrm{Al}}$ is the electronic viscosity of aluminum. The
electronic viscosity can be expressed as $\eta_{\mathrm{Al}}=\left(
2/3\right)  n_{\mathrm{Al}}\tau_{\mathrm{Al}}\left\langle \epsilon
_{\mathrm{Al}}\right\rangle $ where $n_{\mathrm{Al}}$ is the density of free
electrons, $\tau_{\mathrm{Al}}$ is the scattering relaxation time, and
$\left\langle \epsilon_{\mathrm{Al}}\right\rangle $ is the averaged kinetic
energy [see Eq. (43.8) in \cite{Kittel_ESP}]. At low temperatures
$\left\langle \epsilon_{\mathrm{Al}}\right\rangle =3\epsilon_{\mathrm{F,Al}%
}/5$, where $\epsilon_{\mathrm{F,Al}}$ is the Fermi energy. Using the values
$n_{\mathrm{Al}}=18.1\times10^{22}%
%TCIMACRO{\unit{cm}}%
%BeginExpansion
\operatorname{cm}%
%EndExpansion
^{-3}$, $\tau_{\mathrm{Al}}=6.5\times10^{-14}%
%TCIMACRO{\unit{s}}%
%BeginExpansion
\operatorname{s}%
%EndExpansion
$ and $\epsilon_{\mathrm{F,Al}}=11.7%
%TCIMACRO{\unit{eV}}%
%BeginExpansion
\operatorname{eV}%
%EndExpansion
$\ \cite{Ashcroft76} one obtains $\gamma_{\mathrm{N,bulk}}=0.6%
%TCIMACRO{\unit{Hz}}%
%BeginExpansion
\operatorname{Hz}%
%EndExpansion
$. This rough estimate, which disregards confinement, yields a value that is
of the same order of magnitude as the value of $\gamma_{\mathrm{N}}/2\pi=1.0%
%TCIMACRO{\unit{Hz}}%
%BeginExpansion
\operatorname{Hz}%
%EndExpansion
$ that has been obtained above from the fitting of the data with Eq.
(\ref{gamma_m BCS}).

In summary, the contribution of free electrons to the damping rate of an
aluminum mechanical resonator is measured. The behavior near the normal to
super phase transition is compared with a prediction based on the BCS theory.
The relative contribution of free electrons in the normal state to the total
damping rate of the resonator under study is found to be $\gamma_{\mathrm{N}%
}/\left(  \gamma_{\mathrm{N}}+\gamma_{\mathrm{S}}\right)  =0.20$.

This work was supported by the Israel Science Foundation, the Binational
Science Foundation, the Security Research Foundation at Technion and the
Russell Berrie Nanotechnology Institute.

%Just because of unusual number of tables stacked at end
\bibliographystyle{ieeepes}
\bibliography{acompat,Eyal_Bib}
%Produces the bibliography via BibTeX.
\end{document}